# Hybrid-plasticity Photonic Synapses Enabling Hardware-Level Neural Reuse

Chenlei Li[+], Tao Shu[+], Cunyu Shi[1], Wei Wang[2], Shengjie Tang[3], Yueyang Zhang[1], Wei Chen[1], Jungan Wang[3], Bin Li[3], Yu Han[4], Gong Zhang[1], Huan Li[1], Yaocheng Shi[1], Jianwei Wang[5], Feng Qiu[3,4], Daoxin Dai[1,6]*

[1]State Key Laboratory of Extreme Photonics and Instrumentation, Center for Optical & Electromagnetic Research, College of Optical Science and Engineering, International Research Center for Advanced Photonics, Zhejiang University, Zijingang Campus, Hangzhou 310058, China.

[2]Zhejiang University-University of Illinois Urbana-Champaign Institute, Zhejiang University, Haining, China.

[3]Hangzhou Institute for Advanced Study, University of Chinese Academy of Sciences, Hangzhou 10024, Zhejiang Province, China.

[4]Juhe Electro-optic (Hangzhou) Tech. Co. Ltd., Hangzhou 310024, Zhejiang Province, China.

[5]State Key Laboratory for Mesoscopic Physics, School of Physics, Peking University, Beijing, China.

[6]China Jiliang University, Hangzhou 310018, China.

**Corresponding Author: dxdai@zju.edu.cn*

[+] These authors contributed equally to the article

## Abstract

Biological intelligence is distinguished by neural reuse—the capacity to preserve established learning memory while repurposing it for new tasks and dynamic environments. Bringing this capability to photonic hardware requires hybrid plasticity, namely the coexistence of long-term synaptic plasticity for persistent weight storage and short-term synaptic plasticity for rapid, reversible adaptation within a single synaptic element; however, current photonic architectures lack such a unified mechanism. Here, we demonstrate a hybrid-plasticity photonic synapse on thin-film lead zirconate titanate (PZT) that couples non-volatile and volatile modes to enable hardware-level neural reuse. Crucially, high-speed refresh operations can be superimposed without perturbing the stored weight. Such a neural-reuse framework yields a convergence speedup of over 20-fold and reduces the weight updates by approximately 30-fold compared with random initialization. These results establish hybrid-plasticity photonic synapses as a pathway toward on-chip learning systems that are both memory-preserving and rapidly adaptable.

## Introduction

The capacity to learn from past experience and adapt rapidly to new situations —a capability known as *neural reuse*(*1*) is a defining feature of human intelligence. In biological neural circuits, this capability is primarily mediated by synapses, which transmit signals between neurons while storing experience-dependent changes(*2*). Because synapses outnumber neurons by several orders of magnitude(*3*), their plasticity and efficiency largely determine the performance, adaptability, and scalability of neuromorphic systems. Crucially, synaptic function is not limited to the storage of static weights. Instead, neural reuse relies on two complementary forms of synaptic plasticity: long-term plasticity, which encodes durable experience-dependent weight changes, and short-term plasticity, which transiently and reversibly modulates synaptic responses for rapid adaptation to new inputs or tasks. Here, we define hybrid plasticity as the coexistence and cooperative operation of these two forms of synaptic plasticity within a single synapse. As artificial intelligence (AI) advances, imparting hardware with analogous capabilities has become increasingly important, motivating biologically inspired computing paradigms that can retain prior states, adapt efficiently, and reduce the cost of learning new tasks(*4*), as shown in Fig. 1A.

In recent years, electronic synapses based on emerging memory technologies have been extensively explored to alleviate the von Neumann bottleneck and implement synaptic functionality in hardware (*5-7*), including devices based on phase-change chalcogenides, metal oxides, and ferroelectric materials, as well as two-dimensional (2D) layered materials (*7*). These devices have established important routes toward long-term plasticity through non-volatile weight storage and programmable synaptic updates. However, most do not provide a physically integrated short-term plasticity branch that can apply rapid, reversible updates on top of stored weights. This absence of cooperative long-term and short-term plasticity limits their ability to support hardware-level neural reuse. More recently, a ferroionic $CuInP_2S_6$ memristor demonstrated neural reuse by dynamically allocating ferroelectric and ionic phases within a single device(*6*). However, relying on $Cu^+$ ion migration for refresh leads to subsecond-to-second-level update dynamics and limited wafer-scale integrability in vertically stacked van der Waals devices.

Alternatively, photonic neural networks (PNNs) implemented on photonic integrated circuits offer a compelling option, leveraging the inherent parallelism and bandwidth of light to execute high-dimensional matrix operations with ultra-low latency(*8*). Substantial progress has been made in on-chip photonic synapses based on non-volatile photonic memory materials, particularly chalcogenide phase-change materials, which can provide multilevel weight retention with zero static holding power(*9-14*). However, the thermally driven writing processes of phase-change materials are often stochastic, energy-intensive, and susceptible to accumulated variability, while their proximity to the optical mode in waveguides can introduce substantial optical loss, hindering the dynamic adaptation required for complex, evolving tasks in large-scale PNNs(*15*). More fundamentally, unlike biological networks, which preserve learned connections while redeploying them for new functions, existing photonic hardware generally requires stored weights to be fully overwritten when switching tasks(*13*). This rigidity imposes penalties in training latency and energy consumption and motivates a photonic form of hybrid plasticity to provide the device-level basis for hardware-level neural reuse.

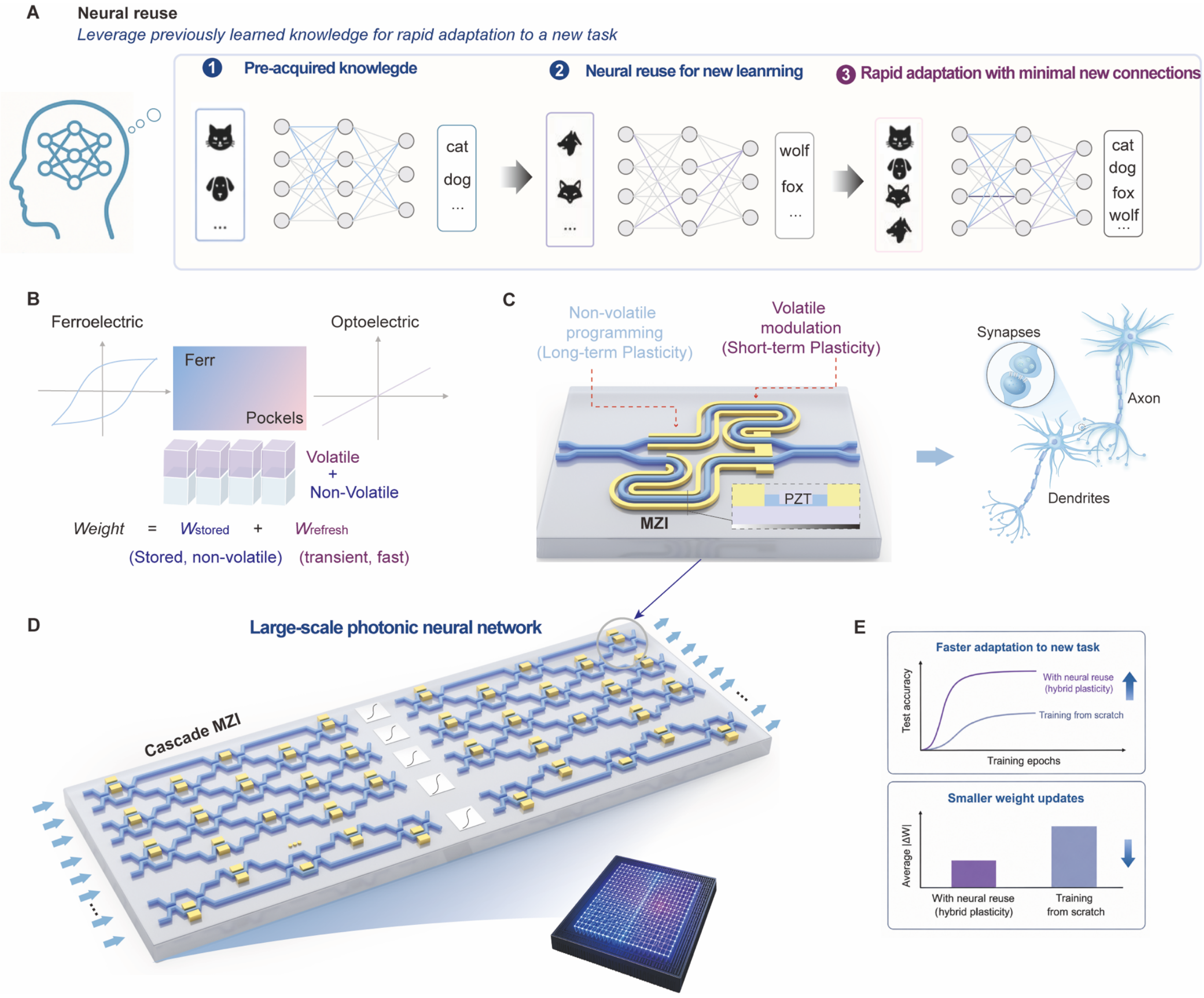

**Fig. 1. Neural reuse concept.** (**A**) Concept of neural reuse for rapid adaptation to new tasks by reusing previously acquired knowledge. (**B**) Coexistence of long-term and short-term plasticity in thin-film PZT: non-volatile ferroelectric programming implements long-term plasticity, whereas volatile Pockels modulation implements short-term plasticity. (**C**) PZT photonic synapse and its analogy to biological synaptic plasticity. (**D**) Large-scale photonic neural-network (PNN) architecture based on cascaded-MZI synapses. (**E**) Neural reuse improves learning efficiency by enabling faster adaptation and smaller weight updates.

Here, we demonstrate a hybrid-plasticity photonic synapse based on thin-film lead zirconate titanate (PZT), in which long-term and short-term synaptic plasticity are implemented through the threshold-governed dual response of the same material(*16*). Above-threshold electrical excitation induces ferroelectric domain switching, producing non-volatile refractive-index changes that persist after the stimulus is removed. This mechanism provides the photonic analogue of long-term plasticity and enables persistent weight storage. By contrast, sub-threshold electrical excitation activates the linear electro-optic Pockels effect without switching ferroelectric domains, producing volatile and fully reversible refractive-index modulation. This mechanism provides the photonic analogue of short-term plasticity and enables rapid synaptic refresh. The coexistence of these two regimes allows task-specific updates to be superimposed on previously learned weights without erasing them, thereby providing a device-level foundation for hardware-level neural reuse (Fig. 1B). The proposed synapse is implemented using a Mach-Zehnder interferometer (MZI) (Fig. 1C and D). We achieve over 5-bit quasi-linear non-volatile weight tuning and experimentally emulate long-term potentiation and depression behaviors, demonstrating stable and repeatable long-term plasticity. The refresh operation is electrically controlled and highly efficient, achieving an ultrafast response time of 35 ps, enabling high-speed short-term plasticity for real-time refresh. We further validate its functional capability through high-speed weight mapping, coherent photonic-network operations, and device-aware simulations based on a residual neural network. This neural reuse architecture accelerates model convergence by more than 20-fold and reduces the required weight updates by approximately 30×, compared with random initialization (Fig. 1E). Furthermore, such a hybrid-plasticity photonic synapse is realized within a single PZT device using only one etching step, which is inherently compatible with wafer-scale fabrication. These results establish hybrid-plasticity PZT photonic synapses as a scalable route toward memory-preserving, rapidly adaptable, and energy-efficient photonic neural networks.

## Photonic Synapses with Hybrid Plasticity

The proposed photonic synapse was fabricated on a 4-inch thin-film PZT-on-$SiO_2$ wafer (see Section I of the supplementary materials for details). To realize hybrid plasticity in a single photonic synapse, the element is designed to support two complementary forms of synaptic plasticity: long-term synaptic plasticity for persistent weight storage and short-term synaptic plasticity for rapid, reversible weight modulation. The present photonic synapse is implemented as an unbalanced MZI, in which electrically induced interference-fringe shifts are converted into optical intensity responses (Section II of the supplementary materials). The near-isotropic electro-optic response of thin-film PZT permits flexible waveguide routing and therefore enables a compact serpentine layout (*17*), reducing the device footprint without imposing stringent crystallographic alignment constraints (Fig. 1C). The same electrode pair is used either to apply high-voltage programming pulses for ferroelectric switching or to deliver low-voltage, high-speed signals for Pockels modulation.

The hybrid plasticity of the PZT photonic synapse originates from the threshold-governed material response that separates persistent ferroelectric programming from transient electro-optic modulation. As illustrated in **Fig. 2**A, when no voltage is applied, the ferroelectric domain configuration remains fixed, and the refractive index retains its initial value. For sub-threshold electrical excitation ($V < V_{th}$), the domains are not switched; instead, the refractive-index is modulated by following the applied electric field through the ultrafast linear Pockels effect. This volatile and fully reversible process provides the photonic analog of short-term synaptic plasticity, enabling rapid synaptic refresh around a stored state. In contrast, when the applied voltage exceeds the switching threshold ($V > V_{th}$), ferroelectric domains reorient toward a new remanent polarization state, producing a non-volatile refractive-index variation that persists after the stimulus is removed. This above-threshold ferroelectric switching provides the photonic analog of long-term synaptic plasticity, enabling persistent weight storage. More details about the underlying switching mechanism are provided in supplementary materials section III.

Here, we treat the output intensity response as the photonic counterpart of the biological postsynaptic potential (PSP) (*18*), and refer to it as the optical postsynaptic potential (OPSP). In particular, by monitoring the wavelength shift of the interference fringes in an unbalanced MZI, we extract the PZT refractive-index change $\Delta n_{PZT}$, which serves as an optical readout of the programmed ferroelectric state (see supplementary materials section II for more details), as shown in **Fig. 2**B.We next quantify these two forms of plasticity using the MZI- based photonc synapses, first by characterizing non-volatile ferroelectric programming as long-term synaptic plasticity, including multilevel weight updates and retention, and then by characterizing the high-speed volatile Pockels response as short-term synaptic plasticity for reversible synaptic refresh.

**Long-term synaptic plasticity**

Long-term plasticity is crucial for neural reuse so that the pretrained weights can remain stable during subsequent adaptation. We therefore first examined the ferroelectric polarization regime, where non-volatile programming stores synaptic weights as persistent phase states, providing the photonic analog of long-term plasticity. To establish deterministic control over this long-term memory, we systematically examine how the amplitude, pulse duration, and pulse count shape the ferroelectric domain evolution. These pulse degrees of freedom provide a route to deterministic long-term plasticity, a key requirement for high-precision, energy-efficient photonic synapses. As depicted in **Fig. 2**C, increasing the pulse amplitude quantitatively controls the switching of ferroelectric domains, which in turn increases the refractive index modulation $\Delta n_{PZT}$. Here, 10 pulses were grouped, with the setting voltage ($V_{set}$) amplitude ranging from 10 V to 38 V. Each pulse had a duration of 100 μs, 500 μs, or 900 μs, while the period was fixed at 1000 μs. Across different pulse groups, a series of 1000 pulses at +100 V was applied to initialize and reset the ferroelectric domains. Crucially, the results in **Fig. 2**C reveal that longer pulse durations yield more polarization switching, as indicated by the comparison among the cases with different durations of 100 μs, 500 μs, and 900 μs, confirming that the change in ferroelectric domain population depends on both amplitude and duration of pulses. It indicates that amplitude and duration act as independent parameters for controlling.

Furthermore, we investigated cumulative switching behavior by applying consecutive 500-μs amplitude-fixed pulses without any intermediate resetting step (**Fig. 2**D). The fraction of switched domains increases monotonically with pulse count ($N$), resulting in synaptic spike-number-dependent plasticity (SNDP), which emulates artificial synaptic plasticity characteristics(*19*). The observation that the proportion of newly switched domains gradually decreases with successive pulses is attributed to the presence of depolarization fields that counter the external field, wherein domains with a lower coercive field switch preferentially. Notably, domain evolution is kept even after $10^6$ pulses. Thus, in addition to the amplitude and duration of pulses, the number of repetitive pulses also serves as a third critical parameter governing the state of synaptic weight, enabling precise control over ferroelectric domain polarization.

The dependence of ferroelectric switching on pulse amplitude, duration, and repetition can be quantitatively modeled, enabling tailored optimization for specific applications. By applying a sequence of incremental voltage pulses under the incremental step pulse programming (ISPP) protocol(*20*), the target state is gradually approached through the controlled domain reorientation without overshoot. As shown in **Fig. 2**E, this protocol achieves quasi-continuous, near-linear 5-bit synaptic weight tuning, establishing a stable and energy-efficient route to deterministic long-term plasticity in programmable photonic systems.

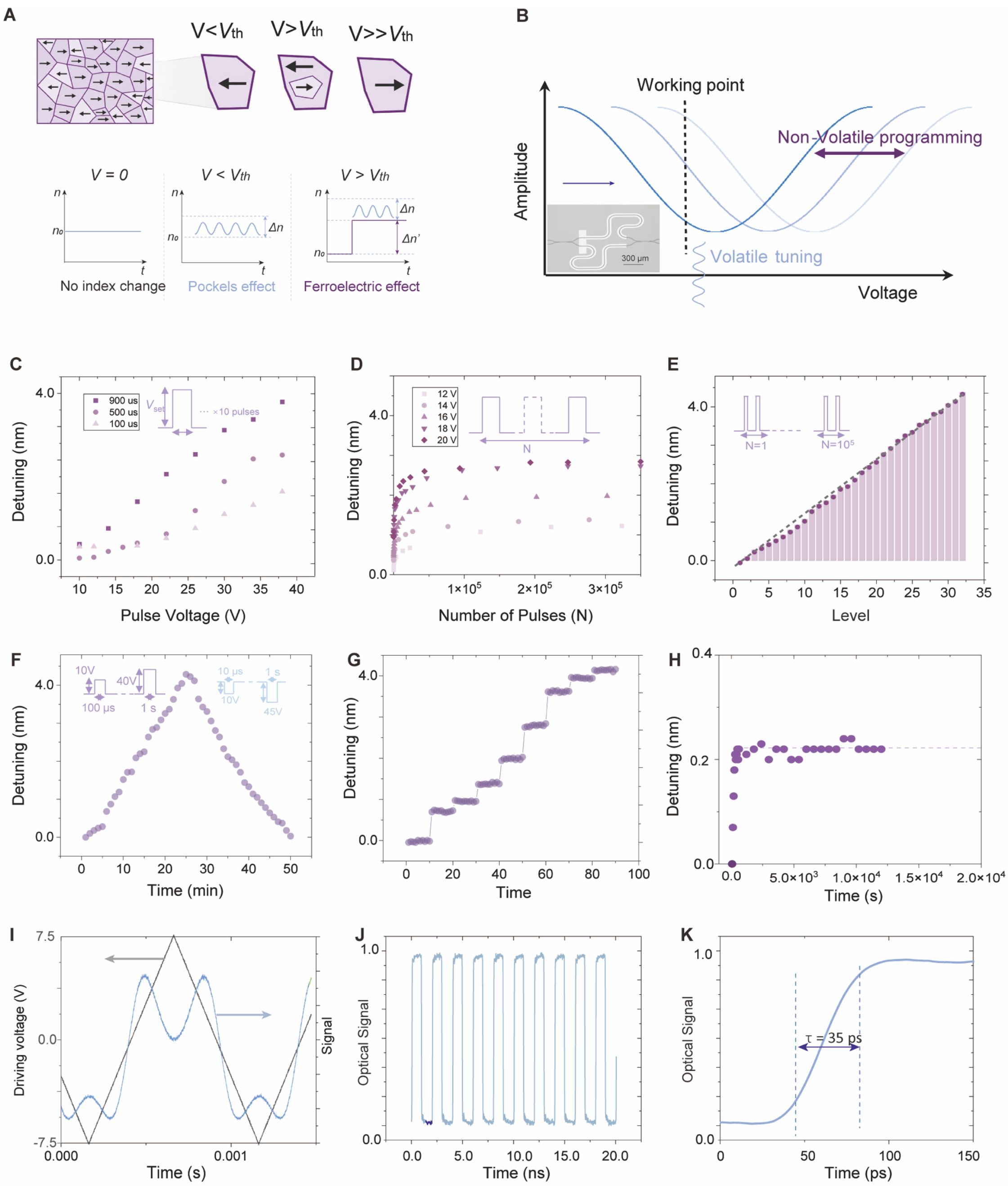

**Fig. 2 Device-level implementation of hybrid plasticity in the PZT photonic synapse.** (**A**) Dual modulation mechanisms in thin-film PZT. Above-threshold excitation writes long-term plasticity through ferroelectric domain switching, whereas sub-threshold excitation produces short-term plasticity through reversible Pockels modulation. (**B**) Experimental protocol for hybrid-plastic operation in the MZI photonic synapse. Above-threshold programming defines the remanent ferroelectric state and optical operating point for long-term weight storage; sub-threshold driving then produces reversible Pockels modulation around this stored state for short-term synaptic refresh, without erasing the non-volatile baseline. Inset: optical micrograph of an MZI-based photonic synapse. (**C**) Fringe shift, proportional to the programmed effective refractive-index change, as a function of programming voltage and pulse duration. (**D**) Cumulative ferroelectric switching under consecutive 500-μs amplitude-fixed pulses without intermediate reset. (**E**) Quasi-continuous 5-bit synaptic weight tuning achieved by ISPP protocol. (**F**) Bidirectional optical postsynaptic potential updates resembling long-term potentiation and long-term depression under programming pulses of opposite polarities. (**G**)

Repeatability of multistate programming measured over nine independent experiments, each repeated 10 times. **(H)** Retention of a programmed non-volatile state over more than 3 hours. (**I**) Modulation efficiency of the photonic synapses, yielding $V_{\pi}L \approx 1.22$ V·cm after initial poling. (**J**) High-speed optical switching waveform driven by sub-threshold Pockels modulation. **(K)** Measured 10%–90% rise time of 35 ps for the volatile electro-optic response.

Biological synapses exhibit bidirectional long-term plasticity, typically manifested as long-term potentiation (LTP) (*21*) and long-term depression (LTD) (*22*)， which were first established experimentally in the hippocampus and cerebellum, respectively. To emulate this bidirectional long-term synaptic plasticity in the PZT photonic synapse, we applied programming pulses with opposite polarities, which drive ferroelectric domains toward distinct polarization states and translate into potentiation- and depression-like OPSP updates. In this experiment, we applied 16 setting pulse groups with voltages ranging from +10 V for a single cycle to + 40 V for 10,000 cycles. We then applied 16 resetting pulse groups with voltages ranging from −10 V for a single cycle to −45 V for 10,000 cycles. Under a sequence of carefully designed voltage pulses, the OPSP can, for the first time, emulate LTP and LTD, while offering a broad continuum of tunable intermediate states (**Fig. 2**F). Importantly, each programmed state exhibits robust non-volatility, enabling stable long-term weight retention across the entire tuning range.

To assess the effectiveness and repeatability of this protocol, we conducted nine independent experiments under identical conditions with different applied voltages. Each experiment was repeated 10 times, with each cycle lasting 1 minute, as shown in Fig. 2G (see supplementary materials, section IV, for details). The non-overlapping across multistates confirms that all programmed states are clearly distinguishable. With further optimization of the applied voltage and pulse width, we anticipate achieving even finer trimming resolution and more resolvable state levels. Finally, we assessed the temporal stability of the non-volatile states by tracking their optical response for three hours after programming (**Fig. 2**H). The programmed state approaches 90% and 99% of its final value within approximately four and five minutes, respectively, indicating rapid stabilization of the ferroelectric domains. No measurable relaxation or reversal was observed over three hours of continuous monitoring at room temperature, signifying that the written ferroelectric configuration is effectively locked. These results confirm that the PZT synapse supports repeatable, bidirectional, and stable long-term plasticity, providing the persistent weight-storage branch required for hybrid plasticity.

**Short-term synaptic plasticity**

While long-term plasticity provides persistent weight storage, refreshable adaptation to downstream tasks requires a complementary mechanism that can transiently modulate synaptic weights rapidly and reversibly without perturbing the stored state. We therefore examined the sub-threshold Pockels-effect regime of PZT as the photonic analog of short-term synaptic plasticity. In this regime, the ferroelectric domains remain unswitched, while the refractive-index change follows the instantaneous electric field without remanence. This volatile and fully reversible response enables ultrafast synaptic refresh around a programmed long-term state.

To quantify this short-term synaptic plasticity, we first characterized its electro-optic efficiency. Following the initial poling, the MZI exhibits a half-wave voltage–length product ($V_{\pi}L$) of approximately 1.22 V·cm (**Fig. 2**I), corresponding to an exceptionally high modulation efficiency compared with state-of-the-art thin-film lithium niobate, while maintaining a compact footprint. To evaluate the dependence of electro-optic performance on the ferroelectric preset state, we further measured the modulation efficiency after programming the device into different non-volatile states. The extracted $V_{\pi}L$ values vary with 1.22, 1.12, 1.02, 1.50, 1.87, and 2.40 V·cm, respectively, when preset at 0, 15, 20, 25, 30, and 60 V for ten minutes—sufficient to reach near-saturation at each voltage level(see supplementary materials section V for details). These results indicate that short-term plasticity can be effectively superimposed on long-term plasticity. The high-speed response of the serpentine-configured MZI was characterized as shown in **Fig. 2**J-K(supplementary materials section VI). The 10%-to-90% rising time ($\tau_{rise}$) and 90%-to-10% falling time ($\tau_{fall}$) are measured at 35 ps and 34 ps, respectively, confirming the ultrafast speed of electro-optic modulation intrinsic to the Pockels effect in PZT. Such picosecond-scale short-term plasticity provides a rapid refresh branch that complements the persistent weight-storage branch established by ferroelectric programming. Together, these results demonstrate that the PZT photonic synapse can support fast, reversible modulation on top of stored long-term weights, providing the essential short-term component of hybrid plasticity.

## Photonic synapses for hardware-level neural reuse

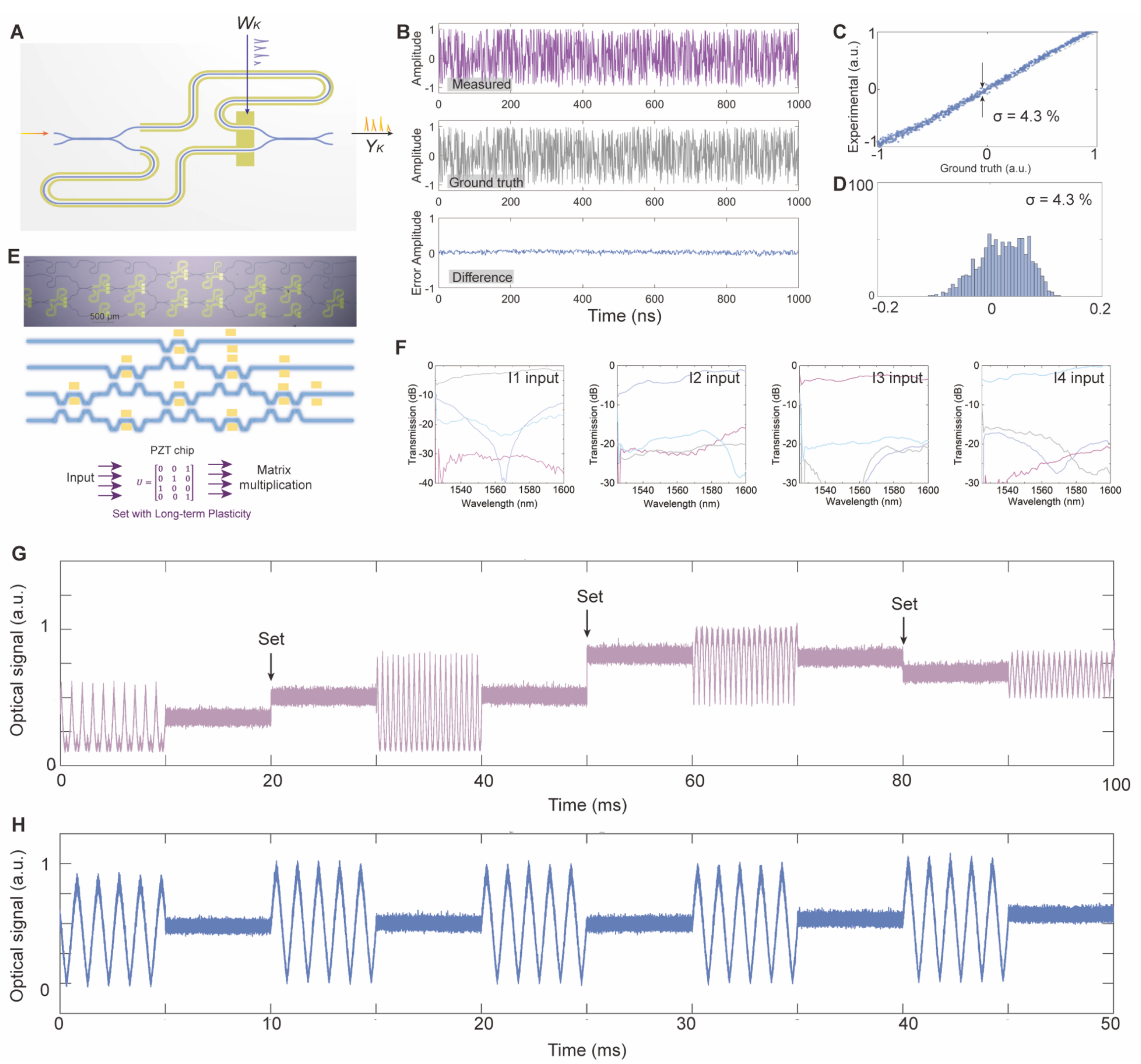

**Fig. 3. Synapse-to-network operation.** (**A**) Schematic of an MZI-based photonic synapse, in which the programmed phase state encodes the synaptic weight $W_k$, and the optical output $Y_k$ represents the corresponding optical postsynaptic response. **(B)** Temporal waveform generated by the photonic synapse. The experimentally measured output is compared with the target ground-truth waveform, with the residual error shown in the lower panel. (**C**) Correlation between the experimentally measured output and the ground-truth signal. The extracted standard deviation of the error is σ=4.3%. (**D)** Statistical distribution of the residual error between the measured and ground-truth waveforms, yielding σ=4.3%. (**E**)Schematic and optical micrograph of a neural network used to implement coherent linear transformations. (**F**) Demonstration of a programmed

4 × 4 column-exchange operation. The target unitary matrix swaps input channels 1 and 3 while preserving channel 2. (**G)** Repeated set-and-refresh operation showing short-term Pockels modulation around different long-term non-volatile baselines. (**H)** Reproducibility of hybrid-plastic operation under repeated alternation between long-term and short-term plasticity.

To establish its role as a photonic computing primitive, we next tested whether the MZI synapse could perform high-speed weight mapping by accurately converting programmed electrical states into optical output levels. As shown in **Fig. 3**A, the PZT photonic synapse encodes the synaptic weight $W_k$, while the MZI converts the corresponding phase state into an optical postsynaptic response $Y_k$. Under sub-threshold excitation, the synapse operates through the linear Pockels effect, enabling reversible and high-speed optical weight mapping without changing the stored ferroelectric state. We programmed the device to reproduce a target temporal waveform and compared the measured optical output with the ground-truth signal. As shown in **Fig. 3**B and **Fig. 3C**, because of the high linearity of the Pockels effect, the measured trace closely follows the target waveform, achieving good consistency without calibration. A point-by-point comparison between experiment and ground truth shows a nearly linear correlation across the full normalized amplitude range, with an error standard deviation of 4.3% at 25 MS/s. The corresponding residual distribution is centered near zero and yields the same standard deviation, confirming that the PZT MZI synapse can provide accurate and reproducible optical weight mapping (Fig. 3D).

Beyond single-synapse operation, practical neural processing requires programmable synaptic elements to be integrated into network-level photonic circuits capable of performing collective linear transformations. Coherent PNNs perform matrix operations through programmable interference in reconfigurable MZI meshes, making such architectures a central hardware platform for high-speed, high-bandwidth, and highly parallel optical computing(*23*). We therefore examined whether PZT synapses could serve as reconfigurable building blocks in a coherent PNN architecture. To this end, we fabricated a programmable 4 × 4 unitary circuit following the Reck scheme(*24*), comprising six cascaded MZIs and six phase shifters. This architecture supports a range of 4 × 4 linear transformations, as illustrated in **Fig. 3E**. In this circuit, non-volatile ferroelectric programming is used to set the optical phases and define the target transfer matrix, thereby providing a persistent hardware configuration for matrix operations. As an example, we implemented a 4×4 column-exchange operation, in which input channels 1 and 3 are swapped while the other channels remain unchanged. **Fig. 3F** shows the measured transmission spectra for all switching states of the programmed unitary circuit, together with the corresponding linear transformation matrix after non-volatile tuning (See supplementary materials section VI for the initial transmission ).

Having separately characterized long-term and short-term plasticity, we next verify their cooperative operation during repeated set-and-refresh cycles. In this regime, a programmed synaptic state serves as a non-volatile baseline, while reversible task-specific updates are transiently applied and removed as required for hardware-level neural reuse. We designed an experimental protocol that combines ferroelectric programming with electro-optic refresh. The device was first initialized to a well-defined remanent state through a programmed electrical pulse sequence (e.g.,100 V, 1000 cycles). In Fig. 3G, above-threshold “Set” operations establish different non-volatile optical baselines, corresponding to distinct remanent ferroelectric states. Around each programmed baseline, sub-threshold triangular driving induces volatile Pockels modulation, which can be repeatedly added and removed without overwriting the stored state. After each transient modulation window, the optical signal returns to the programmed baseline, confirming that short-term refresh can be superimposed on long-term weight storage without mutual interference. We further verified the reproducibility of this behavior, as shown in Fig. 3H, where the optical signal was encoded by the polarization that remains stable after repeated alternation between non-volatile tuning and volatile modulation. These results show that long-term weight retention and short-term dynamic refresh can be superimposed without mutual interference.Together, these results establish that the hybrid plasticity of the PZT-based photonic synapses supports accurate weight programming, network-level linear transformation, and interference-free reversible synaptic refresh on top of persistent weight storage, establishing a device and circuit foundation for hardware-level neural reuse.

### Device–circuit co-simulations of a transfer-learning task

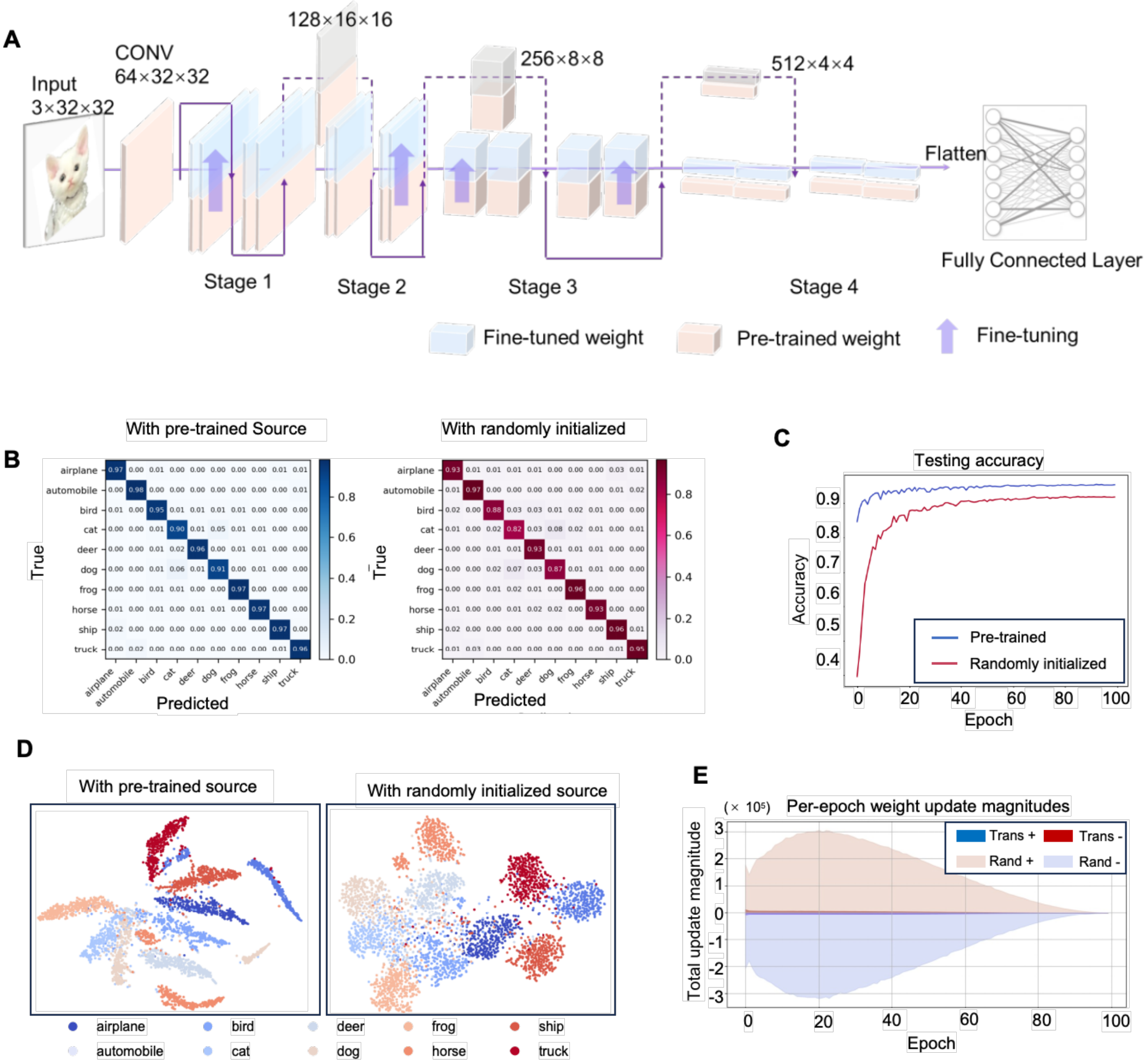

**Fig. 4. Learning performance evaluation.** (**A**) Schematic of the transfer-learning workflow implemented with a residual convolutional neural network. **(B)** Confusion matrices obtained on the CIFAR-10 test set using pretrained initialization and random initialization, showing improved class-wise classification performance enabled by weight reuse. (**C**) Learning curves during model redeployment. The pretrained model reaches the target accuracy with substantially fewer training epochs than the randomly initialized model, indicating faster convergence. (**D**) Two-dimensional visualization of feature distribution based on t-distributed Stochastic Neighbor Embedding (t-SNE) extracted from randomly initialized or pre-trained models on the test set of CIFAR-10. (**E**) Per-epoch weight-update magnitude during CIFAR-10 training, with positive and negative updates plotted above and below zero, respectively.

To further illustrate the functional advantages of hybrid plasticity in neuromorphic contexts, we performed device–circuit co-simulations of a transfer-learning task emulating the neural reuse phenomenon observed in biological systems. Transfer learning enables the application of prior knowledge to new downstream tasks, mirroring the principle of neural reuse and providing a practical framework for assessing our device's potential in adaptive learning(*25*). In this hybrid-plastic framework, non-volatile ferroelectric programming provides persistent storage of pretrained synaptic weights, whereas volatile Pockels modulation enables rapid and reversible weight updates without overwriting the stored state. This division of roles provides the hardware basis for memory-preserving model redeployment across downstream tasks. A residual neural network modeled after ResNet-18 (Fig. 4A) was employed, incorporating the experimentally measured device parameters to evaluate classification performance after model redeployment on the CIFAR-10 dataset. We built a PNN model incorporating the experimentally measured parameters. In the co-simulation, transfer learning proceeds as follows: (1) Pre-training: the weights are initialized and stored in the non-volatile regime to represent the base network trained on dataset A. (2) Task adaptation: The volatile regime is engaged to apply small, reversible weight update for fine-tuning the network to dataset B. (3) Reversion: after adaptation, volatile shifts are removed, restoring the original non-volatile state for reuse in new tasks. Performance metrics, including classification accuracy, convergence speed, and weight-update magnitude, were compared with a randomly initialized baseline.

Leveraging the pre-trained model as an initialization point significantly improved redeployment performance across the evaluated downstream tasks. As shown in Fig. 4B, the pre-trained model achieved consistently higher accuracies across 10 classification categories than random initialization. Upon closer examination of the learning curves (Fig. 4C), it is evident that reusing the pre-trained model resulted in faster learning and superior accuracy. In particular, it reaches 90% accuracy more than 20 times faster than the model trained from random initialization on CIFAR-10. Because the pre-trained model converges in a markedly smaller number of redeployment iterations, it requires a smaller weight-update range and consequently lower power consumption than the model trained from random initialization.

To visualize the advantages of the pre-trained model in classification from the perspective of feature distribution, we utilized the t-distributed Stochastic Neighbor Embedding (t-SNE) algorithm(*26*) to reduce the high-dimensional features extracted by a randomly initialized model or a pre-trained model before the fully connected layer to two-dimensional maps, respectively, and then visualize the categorized feature points accordingly. t-SNE is a nonlinear visualization method that aims to preserve local neighborhood relationships. Each point corresponds to one input sample, positioned according to the feature vector extracted from the network immediately before the final fully connected layer. Therefore, better class separability—manifested as tighter intra-class clusters and clearer inter-class boundaries—indicates that the corresponding model learns more discriminative representations for downstream classification. As shown in Fig. 4D, the pre-trained model learns a markedly more structured and discriminative representation, in which samples from each CIFAR-10 class form compact, well-separated clusters with minimal inter-class overlap. In contrast, the randomly initialized model yields a substantially more entangled embedding, characterized by diffuse class distributions and pronounced mixing among categories. These results indicate that transfer initialization promotes feature separability and representation compactness, consistent with improved classification performance and more stable optimization. We also evaluated the learning curves and the t-SNE–based feature distributions on Oxford-102 Flower for our devices using a pre-trained model and for conventional training with a randomly initialized model (See supplementary materials section VII for details). The results show that the pre-trained model reaches an accuracy of ~80% within 30 epochs, whereas the randomly initialized model remains below 40% even after 100 epochs. Consistent with this gap in learning efficiency, the t-SNE embeddings further reveal that pre-training yields markedly more compact and separable class clusters, while random initialization produces substantially more entangled feature distributions with pronounced inter-class overlap.

We further analyzed the per-epoch magnitudes of the weight updates during CIFAR-10 training, with positive updates plotted above zero and negative updates below. The results are shown in Fig. 4E. We observe that the randomly initialized model undergoes pronounced parameter reconfiguration in the early stage, with update magnitudes rapidly increasing and peaking at around epoch ~20 (reaching $\sim 3\times 10^{5}$), followed by a gradual decay towards near-zero values as the optimization converges. In contrast, the pre-trained model maintains consistently small update magnitudes throughout training, reducing the required weight update by approximately 30× relative to random initialization. This indicates that the parameters begin close to a favorable solution and require only modest fine-tuning. These results underscore the substantially reduced optimization displacement enabled by transfer initialization, consistent with faster stabilization of the learned representation. Collectively, these results demonstrate that neural reuse enables faster learning and reduced resource demands, offering a scalable pathway toward efficient and adaptive photonic–electronic AI systems.

## Discussion and Conclusion

We developed a fully integrated hybrid-plasticity photonic synapse that unifies long-term and short-term synaptic plasticity. Above-threshold ferroelectric switching provides non-volatile long-term plasticity for persistent weight storage, while sub-threshold Pockels modulation provides volatile short-term plasticity for picosecond-scale reversible refresh. This cooperative dual-plasticity mechanism allows rapid task-specific updates to be superimposed on retained weights, thereby enabling hardware-level neural reuse. The synapse provides quasi-continuous 5-bit weight tuning, bidirectional potentiation and depression, repeatable multistate programming, and stable retention, while also supporting picosecond-scale reversible refresh. We further validate its functional capabilities through high-speed weight mapping and coherent photonic network operations. Notably, high-speed refresh can be superimposed without disturbing the stored weight. In device-aware transfer-learning simulations, the resulting neural-reuse framework achieved more than 20-fold faster convergence and approximately 30-fold smaller weight updates than random initialization. These results establish PZT hybrid-plastic synapses as a scalable route toward memory-preserving and rapidly adaptable photonic neural networks.

Beyond the device level, this hybrid-plastic platform also provides a robust pathway toward system-level scalability. The fabrication protocol is potentially compatible with complementary metal-oxide-semiconductor (CMOS) processes (*27*), ensuring wafer-scale photonic integration with exceptional reproducibility. Distinct from those material platforms limited by thermal fragility, the intrinsic thermal robustness of thin-film PZT guarantees reliable operation under elevated temperatures (*28*). Furthermore, PZT serves as a unified foundation for photonic integrated circuits, enabling the monolithic co-integration of high-speed modulation and programmable interferometric processing (*29, 30*). Collectively, these attributes pave the way even for fully integrated optoelectronic systems where lasers(*31*), synapses, neuronal elements (*32*), and photodetectors (*33*) converge on a single chip. Such a platform empowers transfer learning, rapid task redeployment, and sustainable steady-state operation, and establishes a versatile foundation for next-generation photonic hardware capable of practical, hardware-level neural reuse.

**Acknowledgements**

We acknowledge support from the National Natural Science Foundation of China (62550180, 62405271, U23B2047, 62321166651), Zhejiang Provincial Major Research and Development Program (LD19F050001), Fundamental and Interdisciplinary Disciplines Breakthrough Plan of the Ministry of Education of China (JYB2025XDXM106), and Fundamental Research Funds for the Central Universities(226202400171). We also thank Tingbiao Guo for insightful discussions.

**Author contribution.** D.D. and C.L. conceived the idea; C.L. designed the PZT photonic synapses and the chip layout; S.T., T.S., and C.S. fabricated the chip; C.L. performed the transfer learning calculations. T.S., Y.Z., and C.L. performed the high-speed optical signal-processing of the photonic synapses. C.L. and D.D. carried out the data analysis. C.L. and D.D. wrote the manuscript, with input from all the authors. All authors discussed the results and contributed to the manuscript. D.D. supervised the project.

**Competing interests:**

The authors declare no competing interests.